\documentclass[%
 reprint,
 amsmath,amssymb,
 aps,
]{revtex4-2}

\usepackage{graphicx}
\usepackage{dcolumn}
\usepackage{bm}
\usepackage{subfigure}
\usepackage[dvipsnames]{xcolor}
\usepackage{amssymb,mathtools}
\usepackage{bigints}

\begin{document}

\preprint{APS/123-QED}

\title{On the Experimental Determination of Nonlocal Characteristics of Two-Qubit Gates}

\author{M. Karthick Selvan}
\email{karthick.selvan@yahoo.com}%

\author{S. Balakrishnan}%
\email{physicsbalki@gmail.com}
\affiliation{Department of Physics, School of Advanced Sciences, Vellore Institute of Technology, Vellore - 632014, Tamilnadu, India.}%



\begin{abstract}
In this paper, we discuss the experimental determination of the nonlocal characteristics of two-qubit gates. Based on the recently derived expressions for the entangling power and gate typicality of two-qubit gates, we construct two-qubit quantum circuits to measure the entangling power and gate typicality of the two-qubit gates, which are generated by the elements of the $su(4)$ Cartan subalgebra. These elements describe the native interactions in many quantum processors. Hence, these circuits can be used to determine the nonlocal characteristics of native gates of many quantum processors. In each circuit, the native gate is applied twice. In addition, each circuit consists of at least one CNOT gate. A set of six two-qubit circuits is constructed to measure the entangling power. The number of circuits is further reduced to three by increasing the nonlocal resources (the number of CNOT gates). To measure the gate typicality, a set of three two-qubit circuits is constructed. Measurement of gate typicality requires more nonlocal resources than the measurement of entangling power.
\end{abstract}

\maketitle

\section{Introduction}
Nonlocal two-qubit gates can create entangled states and transfer the quantum information between different qubits. They are used in the universal gate set to perform quantum computation~\cite{Barenco1995,DiVincenzo1995}. Hence, it is essential to study the experimental characterization of the nonlocal properties of two-qubit gates. There exist measures to quantify the nonlocal characteristics of two-qubit gates. Among them, the linear entropy~\cite{Zanardi2001}, which quantifies operator entanglement, and the entangling power~\cite{Zanardi2000}, measuring the ability of two-qubit gates to generate entangled states, have been studied extensively. Four-qubit (or four-qudit) protocols to determine the linear entropy and entangling power of two-qubit (or two-qudit) unitary operations have been proposed~\cite{Zanardi2000,Zanardi2001}. The gate typicality~\cite{Jonnadula2017,Jonnadula2020}, a local invariant complementary to entangling power, has been studied.  

There exist techniques to characterize the implementation of two-qubit gates~\cite{Tripathi2025}. Entanglement of the states generated by the two-qubit gates has been measured~\cite{Selvan2024}. Analytical studies on the nonlocal fidelity and the entangling power difference between the ideal cross-resonance gate~\cite{Rigetti2010} and the echoed cross-resonance gate have been carried out~\cite{Malekakhlagh2020}. The Cartan coordinates and entangling power of cross-resonance pulse gate were experimentally estimated using unitary process tomography (UPT)~\cite{Sugawara2025}. In UPT, 36 two-qubit quantum circuits need to be executed~\cite{Sugawara2025}. A method to investigate the crosstalk between qubits during the implementation of a perfect entangler was recently proposed~\cite{Krauss2026}. It involves plotting a quantity, which consists of two perfect entangler functionals~\cite{Goerz2015} corresponding to two different spectator qubit states and the weighted infidelity between the perfect entanglers implemented with the spectator qubit at two different states, as a function of the spectator qubit frequency. 

Recently, the entangling power, gate typicality, and the linear entropy of a two-qubit gate are expressed in terms of the squared length of the chords in the Argand diagram of the squared eigenvalues of the nonlocal part of the two-qubit gate~\cite{Selvan2026}. The three nonlocal measures are also expressable in terms of the squared perpendicular distances between the chords and the zero. In this paper, we construct two-qubit quantum circuits to measure these distances and quantify the entangling power, gate typicality, and linear entropy of the nonlocal part of two-qubit gates. The nonlocal part of two-qubit gates is generated by the $su(4)$ Cartan subalgebra~\cite{Zhang2003}, and they are natively realizable on many quantum processors~\cite{Debnath2016,Huang2023,Wei2024}. Hence, these circuits can be used to study the nonlocal characteristics of the native gates of many quantum processors. Apart from implementing the native gate of interest, the circuits require additional nonlocal resources. Each circuit consists of unitary matrices that transform the computational basis into the magic basis and its inverse . These unitary matrices belong to the CNOT equivalence class~\cite{Vatan2004}. Hence, a CNOT gate can be used to implement these two unitary matrices. 

This paper is organized as follows. In Section II, the canonical decomposition of two-qubit gates and nonlocal measures of two-qubit gates are briefly described. In Section III, the circuits are constructed to measure the entangling power and gate typicality. In Section IV, the conclusion is provided.

\section{Background}

\subsection{Canonical decomposition of two-qubit gates}
A two-qubit gate, $U \in \text{SU}(4)$, can be decomposed as follows~\cite{Zhang2003}. 
\begin{equation}\label{eq1}
U = (k_1 \otimes k_2) U_d (c_1, c_2, c_3) (k_3 \otimes k_4),
\end{equation}
where $k_{1(3)} \otimes k_{2(4)} \in SU(2) \otimes SU(2)$ are the local parts of $U$, 
\begin{equation}
U_d(c_1, c_2, c_3) = \exp \left[ \dfrac{i}{2} \left( c_1 \sigma_x^{\otimes 2} + c_2 \sigma_y^{\otimes 2} + c_3 \sigma_z ^{\otimes 2} \right) \right] 
\end{equation}
is the nonlocal part of $U$ and $(c_1, c_2, c_3)$ are Cartan co-ordinates. 

\begin{figure}[h]
\includegraphics[width=0.5\textwidth]{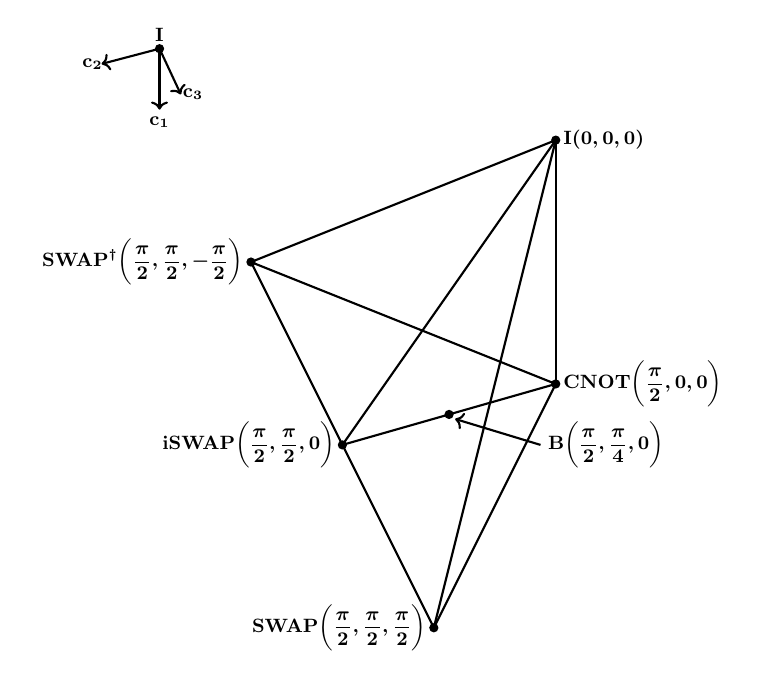}
\caption{Geometry of local equivalence classes of two-qubit gates}
\label{fig0}
\end{figure}

Two-qubit gates that can be obtained from one another by local unitary operations form a local equivalence class, and the local equivalence classes of two-qubit gates are geometrically represented by the points of a tetrahedron shown in FIG.~\ref{fig0}. Each point of this tetrahedron is the Cartan co-ordinates of a local equivalence class of two-qubit gates, and they satisfy the condition: $\pi/2 \geq c_1 \geq c_2 \geq \vert c_3 \vert \geq 0$. 

Local invariants of two-qubit gates, which can be used to distinguish the two-qubit gates belonging to different local equivalence classes, can be written in terms of Cartan co-ordinates as follows~\cite{Watts2013}. 

\begin{align}
g_1 & = \dfrac{1}{4} [\cos(2c_1) + \cos(2c_2) + \cos(2c_3)  \nonumber \\
 & + \cos(2c_1) \cos(2c_2) \cos(2c_3)],  \nonumber \\
g_2 & = \dfrac{1}{4} \left[\sin(2c_1) \sin(2c_2) \sin(2c_3) \right], \nonumber
\\
g_3 & = \cos(2c_1) + \cos(2c_2) + \cos(2c_3).
\end{align}

The nonlocal characteristics of the two-qubit gate $U$ are determined by its nonlocal part $U_d$. The eigenstates of the nonlocal part $U_d$ 
\begin{equation}\label{eqn3.1}
U_d \vert \Psi_k \rangle = e^{ih_k/2} \vert \Psi_k \rangle~~~~~(k=1,2,3,4)
\end{equation}
with 
\begin{equation}\label{eq3.2}
\vert \Psi_1 \rangle = \dfrac{\vert 00 \rangle + \vert 11 \rangle}{\sqrt{2}},~ h_1 = c_1 - c_2 + c_3,
\end{equation}
\begin{equation}\label{eq3.3}
\vert \Psi_2 \rangle = \dfrac{i \vert 01 \rangle + i\vert 10 \rangle}{\sqrt{2}},~ h_2 = c_1 + c_2 - c_3,
\end{equation}
\begin{equation}\label{eq3.4}
\vert \Psi_3 \rangle = \dfrac{ \vert 01 \rangle - \vert 10 \rangle}{\sqrt{2}},~ h_3 = - c_1 - c_2 - c_3,
\end{equation}
and
\begin{equation}\label{eq3.5}
\vert \Psi_4 \rangle = \dfrac{i \vert 00 \rangle - i\vert 11 \rangle}{\sqrt{2}},~h_4 = - c_1 + c_2 + c_3,
\end{equation}
form the magic basis~\cite{Zhang2003,Makhlin2002}. 

The transformation from the computational basis into the magic basis is caused by the following unitary matrix. 
\begin{equation}\label{eq3.6}
Q = \dfrac{1}{\sqrt{2}} \begin{bmatrix} 1 & 0 & 0 & i \\ 0 & i & 1 & 0 \\ 0 & i & -1 & 0 \\ 1 & 0 & 0 & -i \end{bmatrix}. 
\end{equation}

\subsection{Nonlocal characteristics of two-qubit gates}

The operator entanglement and the ability to generate entangled states are two main nonlocal characteristics of two-qubit gates. The operator entanglement of a two-qubit gate $U$ is defined as the bipartite entanglement of the following four-qubit state $\vert \Psi (U) \rangle_{A_1 A_2 B_1 B_2}$ corresponding to the partition $A_1 A_2 \vert B_1 B_2$~\cite{Zanardi2001}. 

\begin{equation}\label{eqn2.2.1}
\vert \Psi (U) \rangle_{A_1 A_2 B_1 B_2} = \left[ U_{A_1 B_1} \otimes I_{A_2 B_2} \right] \left\{ \vert \Psi'_+ \rangle_{A_1 A_2} \otimes \vert \Psi'_+ \rangle_{B_1 B_2} \right\},
\end{equation}
where $\vert \Psi'_+ \rangle = \vert 00 \rangle + \vert 11 \rangle$ is one of the unnormalized maximally entangled Bell states. 

Linear entropy $(L)$ is a measure of operator entanglement of two-qubit gates. For a two-qubit gate $U$, it is defined as 

\begin{equation}\label{eqn2.2.2}
L(U) = 1 - tr \left[ \rho_{A_1 A_2}^2 \right],
\end{equation}
where $\rho_{A_1 A_2} = tr_{B_1 B_2} \left[ \vert \Psi(U) \rangle \langle \Psi(U) \vert_{A_1 A_2 B_1 B_2} \right]$. 

In terms of Cartan coordinates and local invariants of the two-qubit gate $U$, the linear entropy of $U$ is expressed as 
\begin{align}\label{eqn2.2.3}
L(U) & = 1 - \dfrac{1}{4} \bigg[ 1 + \cos^2 (c_1) \cos^2 (c_2) + \cos^2 (c_2) \cos^2 (c_3) \nonumber \\ 
& + \cos^2(c_3) \cos^2(c_1) \bigg],
\end{align}
and 
\begin{equation}\label{eqn2.2.4}
L(U) = 1 - \dfrac{1}{8} \left[ 3 + 2 \sqrt{g_1^2 + g_2^2} + g_3 \right],
\end{equation}
respectively~\cite{Balakrishnan2011}. 

Entangling power is a measure of the ability of two-qubit gates to generate entangled states. For a two-qubit gate $U$, entangling power is defined as the average entanglement generated over all product states by $U$~\cite{Zanardi2000}. 

\begin{equation}\label{eqn2.2.5}
e_p(U) = \overline{L(U \left[ \vert \psi_A \rangle \otimes \vert \psi_B \rangle \right])}^{\vert \psi_A \rangle, \vert \psi_B \rangle},
\end{equation}
where $L$ is the linear entropy of the two-qubit state $U\left[ \vert \psi_A \rangle \otimes \vert \psi_B \rangle \right]$ and the overline denotes the average over all product states distributed according to some probability distribution. 

For uniform distribution of product states, the entangling power of $U$ is expressed in terms of linear entropies of $U$, $US$ (product of $U$ and the SWAP gate $S$), and $S$ (the SWAP gate) as follows~\cite{Zanardi2001}. 

\begin{equation}\label{eqn2.2.6}
e_p(U) = \dfrac{4}{3} \left[ L(U) + L(US) - L(S) \right],
\end{equation}

The entangling power of two-qubit gate $U$ with Cartan coordinates $(c_1, c_2, c_3)$ and local invariants $(g_1, g_2, g_3)$ also have the following expressions~\cite{Rezakhani2004,Balakrishnan2010}. 
\begin{align}\label{eqn2.2.7}
e_p(U) & = \dfrac{1}{18} \big[ 3 - \cos(2c_1) \cos(2c_2)  \nonumber \\ 
& - \cos(2c_2) \cos(2c_3) - \cos(2c_3) \cos(2c_1) \big],
\end{align}
and 
\begin{equation}\label{eqn2.2.8}
e_p(U) = \dfrac{2}{9} \left[ 1 - \sqrt{g_1^2 + g_2^2} \right].
\end{equation}

A local invariant, called gate typicality, was proposed as complementary to entangling power~\cite{Jonnadula2017}. For the two-qubit gate $U$, it is expressed as an antisymmetric combination of $L(U)$ and $L(US)-L(S)$ as follows~\cite{Jonnadula2020}. 

\begin{equation}\label{eqn2.2.9}
g_t(U) = \dfrac{2}{3} \left[ L(U) - L(US) + L(S) \right].
\end{equation}

It is expressed in terms of Cartan coordinates $(c_1, c_2, c_3)$ as~\cite{Jonnadula2020}
\begin{equation}\label{eqn2.2.10}
g_t(U) = \dfrac{1}{3} \left[ \sin^2(c_1) + \sin^2(c_2) + \sin^2(c_3) \right].
\end{equation}

It is also a linear function of the local invariant $g_3$~\cite{Selvan2023}. That is, 
\begin{equation}\label{eqn2.2.11}
g_t(U) = \dfrac{3 - g_3}{6}.
\end{equation}

From Eqns.~\ref{eqn2.2.6} and \ref{eqn2.2.9}, the linear entropy of a two-qubit gate can be expressed in terms of entangling power and gate typicality as follows. 
\begin{equation}\label{eqn2.2.12}
L(U) = \dfrac{3}{4} \left[ \dfrac{3}{2} e_p(U) + g_t(U) \right].
\end{equation}

\section{Experimental determination of Entangling power and gate typicality}
\begin{figure}[h]
\centering
\includegraphics[scale=0.65]{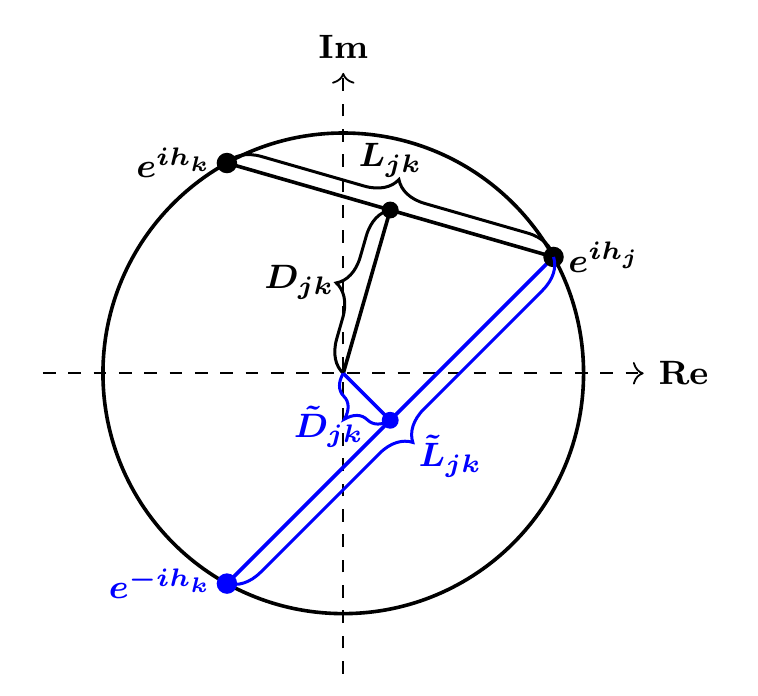}
\caption{Argand diagram showing the chords connecting the squared eigenvalue $e^{ih_j}$ with $e^{\pm ih_k}$ and the perpendicular distances of the chords from the origin.}
\label{fig1}
\end{figure}

It has been shown that the entangling power of a two-qubit gate $U$ with the nonlocal part $U_d$ can be expressed in terms of the squared length of the chords connecting the squared eigenvalues of $U_d$ with each other in the Argand diagram~\cite{Selvan2026}. The expression is given by
\begin{equation}\label{eqn3.1}
e_p(U_d) = \dfrac{1}{18} \sum_{j=1}^3 \sum_{k=j+1}^4 \dfrac{L_{jk}^2}{4},
\end{equation}
where $L_{jk} = \vert e^{ih_j} - e^{ih_k} \vert$ as shown in FIG.~\ref{fig1}. 

From FIG.~\ref{fig1} and Eq.~\ref{eqn3.1}, the entangling power can also be expressed as
\begin{equation}\label{eqn3.2}
e_p(U_d) = \dfrac{1}{18} \left[ 6 - \sum_{j=1}^3 \sum_{k=j+1}^4 D_{jk}^2 \right],
\end{equation}
where $D_{jk} = \dfrac{\vert e^{ih_j} + e^{ih_k} \vert}{2}$. 

It can be verified that 
\begin{equation}\label{eqn3.3}
\dfrac{L_{jk}^2}{4} + D_{jk}^2 = 1. 
\end{equation}

\begin{figure*}
\begin{tabular}{c}
\includegraphics[scale=1.0]{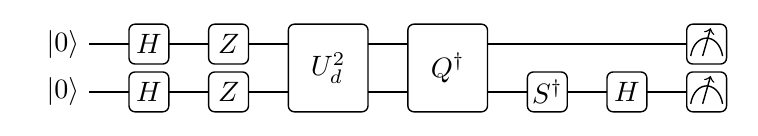} \\ (a) \\
\includegraphics[scale=1.0]{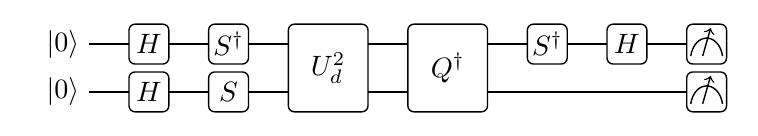} \\ (b) \\
\includegraphics[scale=1.0]{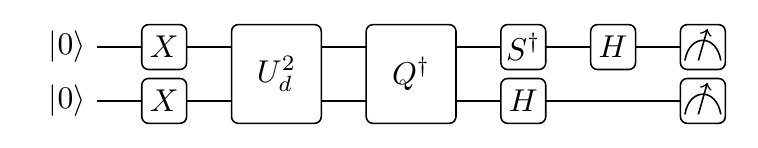} \\ (c) \\
\includegraphics[scale=1.0]{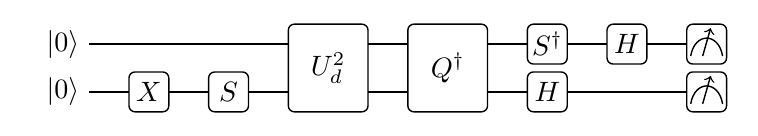} \\ (d) \\
\includegraphics[scale=1.0]{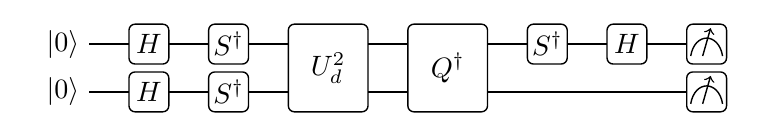} \\ (e) \\
\includegraphics[scale=1.0]{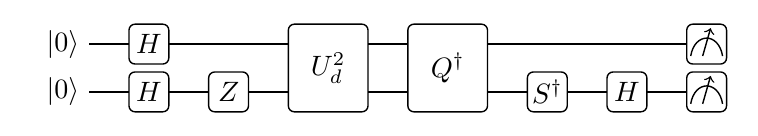} \\ (f) \\
\end{tabular}
\caption{Circuits to measure (a) $(L_{12}^2/4,~D_{12}^2)$, (b) $(L_{13}^2/4,~D_{13}^2)$, (c) $(L_{14}^2/4,~D_{14}^2)$, (d) $(L_{23}^2/4,~D_{23}^2)$, (e) $(L_{24}^2/4,~D_{24}^2)$, and (f) $(L_{34}^2/4,~D_{34}^2)$}
\label{fig3}
\end{figure*}

Both $L_{jk}^2/4$ and $D_{jk}^2$ for $jk \in \{12, 13, 14, 23, 24, 34\}$ can be measured using the circuits shown in FIG.~\ref{fig3}. The circuit shown in FIG.~\ref{fig3}a, can be used to measure $L_{12}^2/4$ and $D_{12}^2$. It is explained in the following. 

\begin{enumerate}
\item The local operations $(H \otimes H)$ and $(Z \otimes Z)$ with $H = \dfrac{1}{\sqrt{2}} \begin{bmatrix}
1 & 1 \\ 1 & -1
\end{bmatrix} $ and $Z=\begin{bmatrix}
1 & 0 \\ 0 & -1 
\end{bmatrix}$ causes the following transformation. 
\begin{equation}
\vert \psi_1 \rangle = \vert 00 \rangle \rightarrow \vert \psi_2 \rangle = \dfrac{\vert \Psi_1 \rangle + i \vert \Psi_2 \rangle}{\sqrt{2}}. 
\end{equation}
\item Application of $U_d$ twice on the state $\vert \psi_2 \rangle$ results in the following transformation. 
\begin{equation}
\vert \psi_2 \rangle \rightarrow \vert \psi_3 \rangle = \dfrac{e^{ih_1} \vert \Psi_1 \rangle + i e^{ih_2} \vert \Psi_2 \rangle}{\sqrt{2}}.
\end{equation}
\item The unitary operation $Q^\dagger$ transforms $\vert \psi_3 \rangle$ as follows.
\begin{equation}
\vert \psi_3 \rangle \rightarrow \vert \psi_4 \rangle = \dfrac{e^{ih_1} \vert 00 \rangle + i e^{ih_2} \vert 01 \rangle}{\sqrt{2}}
\end{equation}
\item Finally, the actions of $S^\dagger = \begin{bmatrix}
1 & 0 \\ 0 & -i 
\end{bmatrix}$ and $H$ on the second qubit (qubits are arranged from left to right) gives the following state. 
\begin{equation}
\vert \psi_5 \rangle = \left( \dfrac{e^{ih_1}+e^{ih_2}}{2} \right) \vert 00 \rangle + \left( \dfrac{e^{ih_1}-e^{ih_2}}{2} \right) \vert 01 \rangle. 
\end{equation}
Thus, $P_{\vert 00 \rangle} = D_{12}^2$ and $P_{\vert 01 \rangle} = L_{12}^2/4$. 
\end{enumerate}

Similarly, the operation of other circuits can also be explained. In the circuits, shown in FIGs.~\ref{fig3}c and \ref{fig3}d, $X= \begin{bmatrix}
0 & 1 \\ 1 & 0
\end{bmatrix}$, and the outputs of the circuits are $\{P_{\vert 00 \rangle} = P_{\vert 11 \rangle} = D_{14}^2/2, P_{\vert 01 \rangle} = P_{\vert 10 \rangle} = L_{14}^2/8\}$ and $\{P_{\vert 00 \rangle} = P_{\vert 11 \rangle} = D_{23}^2/2, P_{\vert 01 \rangle} = P_{\vert 10 \rangle} = L_{23}^2/8\}$, respectively. 

\begin{figure}[h]
\begin{tabular}{c}
\includegraphics[scale=0.55]{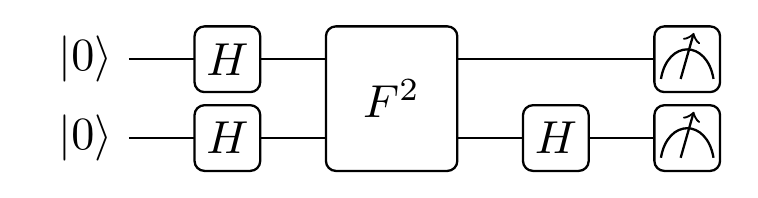} \\ (a) \\
\includegraphics[scale=0.55]{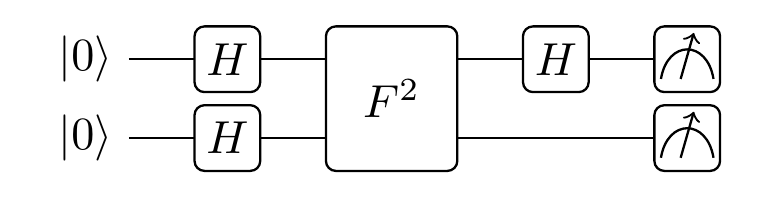} \\ (b) \\
\includegraphics[scale=0.55]{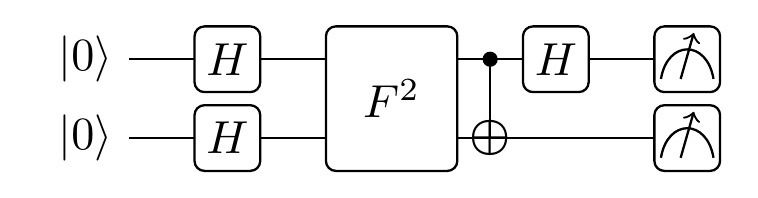} \\ (c) \\
\end{tabular}
\caption{Circuits to measure (a) $(D_{12}^2/2, L_{12}^2/8)$ and $(D_{34}^2/2, L_{34}^2/8)$, (b) $(D_{13}^2/2, L_{13}^2/8)$ and $(D_{24}^2/2, L_{24}^2/8)$, and (c) $(D_{14}^2/2, L_{14}^2/8)$ and $(D_{23}^2/2, L_{23}^2/8)$.}
\label{fig4}
\end{figure}

Each circuit, shown in FIG.~\ref{fig3}, provides one set of complementary quantities $\{D_{jk}^2, L_{jk}^2/4\}$. It is possible to construct a circuit to get two sets of complementary quantities by increasing the nonlocal resources. Such circuits are shown in FIG.~\ref{fig4}, where $F= Q^\dagger U_d Q$. The operation of the first circuit (FIG.~\ref{fig4}a) is explained in the following. 
\begin{enumerate}
\item The initial state $\vert \psi'_1 \rangle = \vert 00 \rangle$ is tranformed into $\vert \psi'_2 \rangle = \dfrac{\vert 00 \rangle + \vert 01 \rangle \rangle + \vert 10 \rangle + \vert 11 \rangle}{2}$ by $H \otimes H$. 
\item The action of $F^2$, where $Q$ acts first followed by $U_d^2$ and $Q^\dagger$, transforms $\vert \psi'_2 \rangle$ into the following state. 
\begin{equation}\label{psi3}
\vert \psi'_3 \rangle = \dfrac{\vert 0 \rangle \left( e^{ih_1} \vert 0 \rangle + e^{ih_2} \vert 1 \rangle \right) + \vert 1 \rangle \left( e^{ih_3} \vert 0 \rangle + e^{ih_4} \vert 1 \rangle \right)}{2}. 
\end{equation}
\item The action of the $H$ on the second qubit transforms $\vert \psi'_3 \rangle$ into 
\begin{align}
\vert \psi'_4 \rangle & = \left(\dfrac{e^{ih_1} + e^{ih_2}}{2\sqrt{2}} \right) \vert 00 \rangle + \left(\dfrac{e^{ih_1} - e^{ih_2}}{2\sqrt{2}} \right) \vert 01 \rangle   \nonumber \\ & + \left(\dfrac{e^{ih_3} + e^{ih_4}}{2\sqrt{2}} \right) \vert 10 \rangle + \left(\dfrac{e^{ih_3} - e^{ih_4}}{2\sqrt{2}} \right) \vert 11 \rangle. 
\end{align}
Thus, $P_{\vert 00 \rangle} = D_{12}^2/2$, $P_{\vert 01 \rangle} = L_{12}^2/8$, $P_{\vert 10 \rangle} = D_{34}^2/2$, and $P_{\vert 11 \rangle} = L_{34}^2/8$. 
\end{enumerate}
The state $\vert \psi'_3 \rangle$ in Eq.~\ref{psi3} can also be written as 
\begin{equation}
\vert \psi'_3 \rangle = \dfrac{ \left( e^{ih_1} \vert 0 \rangle + e^{ih_3} \vert 1 \rangle \right) \vert 0 \rangle + \left( e^{ih_2} \vert 0 \rangle + e^{ih_4} \vert 1 \rangle \right) \vert 1 \rangle }{2}. 
\end{equation} 

Hence, applying $H$ on the first qubit (FIG.~\ref{fig4}b) , the sets ${(D_{13}^2/2, L_{13}^2/8), (D_{24}^2/2, L_{24}^2/8) }$ can be measured. The circuit, shown in FIG.~\ref{fig4}c can be verified to provide $(D_{14}^2/2, L_{14}^2/8)$ and $(D_{23}^2/2, L_{23}^2/8)$. 

Similar to entangling power, the gate typicality can be expressed in terms of the squared length of the chords connecting $e^{ih_1}$ with the complex conjugate of the other three squared eigenvalues~\cite{Selvan2026}. It can also be expressed in terms of the perpendicular distances of the chords from the origin. The expressions are given below. 

\begin{equation}
g_t(U_d) = \dfrac{1}{12} \sum_{k=2}^4 \tilde{L}_{1k}^2 = \dfrac{1}{3} \left[3 - \sum_{k=2}^4 \tilde{D}_{1k}^2 \right],
\end{equation}
where $\tilde{L}_{1k} = \vert e^{ih_1} -e^{-ih_k} \vert$, $\tilde{D}_{1k} = \vert e^{ih_1} + e^{-ih_k} \vert/2$, and $\dfrac{\tilde{L}_{1k}^2}{4} + \tilde{D}_{1k}^2 = 1$.  

\begin{figure*}
\begin{tabular}{c}
\includegraphics[scale=1.0]{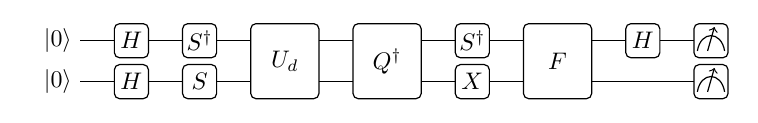} \\ (a) \\
\includegraphics[scale=1.0]{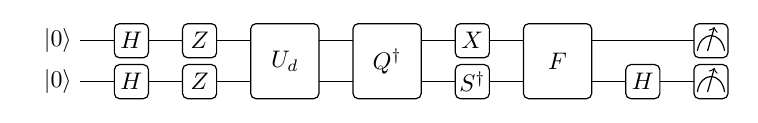} \\ (b) \\
\includegraphics[scale=1.0]{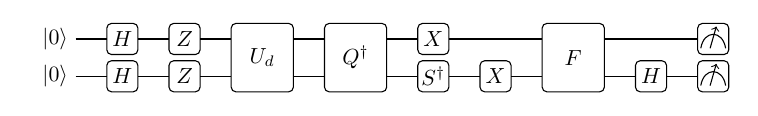} \\ (c) \\
\end{tabular}
\caption{Circuits to measure (a) $(\tilde{D}_{12}^2, \tilde{L}_{12}^2/4)$, (b) $(\tilde{D}_{13}^2, \tilde{L}_{13}^2/4)$, and (d) $(\tilde{D}_{14}^2, \tilde{L}_{14}^2/4)$ }
\label{fig5}
\end{figure*}

Circuits to measure $(\tilde{D}_{1k}^2, \tilde{L}_{1k}^2/4)$ are shown in FIG.~\ref{fig5}. The circuit in FIG.~\ref{fig5}a provides $P_{\vert 01 \rangle} = \tilde{D}_{12}^2$ and $P_{\vert 11 \rangle} = \tilde{L}_{12}^2/4$. The output probabilities for the circuits, shown in FIGs.~\ref{fig5}b and \ref{fig5}c, are $\{P_{\vert 00 \rangle} = P_{\vert 01 \rangle} = 0, P_{\vert 10 \rangle} = \tilde{D}_{13}^2, P_{\vert 11 \rangle} = \tilde{L}_{13}^2/4 \}$ and $\{P_{\vert 00 \rangle} = P_{\vert 01 \rangle} = 0, P_{\vert 10 \rangle} = \tilde{D}_{14}^2, P_{\vert 11 \rangle} = \tilde{L}_{14}^2/4 \}$, respectively. In the derivation of these probabilities, the relation $h_1 + h_2 + h_3 + h_4 = 0$ is utilized. 

\section{Conclusion}

We have constructed circuits to measure the entangling power and gate typicality of the nonlocal part of two-qubit gates. The linear entropy can be obtained by substituting the measured values of entangling power and gate typicality in Eq.~\ref{eqn2.2.12}. To use these circuits for a general two-qubit gate, the knowledge of the canonical decomposition of the two-qubit gate is required. All the circuits require two applications of the two-qubit gate of interest. The first set of six circuits, constructed to measure the entangling power, requires a CNOT gate. The first two circuits in the second set require two CNOT gates, and the last circuit needs three CNOT gates. All three circuits, constructed to measure the gate typicality, use three CNOT gates. 

These circuits can be used in quantum processors to characterize the nonlocal properties of two-qubit evolution, provided the evolution is unitary. In NISQ (noisy intermediate-scale quantum) devices, the evolution is not unitary due to noise. The entangling power and gate typicality expressions, used in this paper, are not valid for non-unitary evolutions. However, since these circuits use only two qubits, have small depth, and involve only measurement of probabilities, they can be used to estimate the nonlocal characteristics of the underlying unitary evolution.


\end{document}